\documentclass{article}%
\usepackage{amsmath}
\usepackage{amsfonts}
\usepackage{amssymb}

\usepackage{color}
\usepackage{graphicx}
\usepackage{pstcol,times,graphicx,graphics,pstricks,pst-node,pst-coil}

\newrgbcolor{c0} {0.84 0.11 0.49}
\newrgbcolor{c1} {0.64 0.11 0.84}
\newrgbcolor{c2} {0.38 0.10 0.84}
\newrgbcolor{c3} {0.08 0.57 0.86}
\newrgbcolor{c4} {0.08 0.90 0.08}
\newrgbcolor{c5} {0.80 0.80 0.20}
\newrgbcolor{c6} {1.00 0.50 0.25}
\newrgbcolor{c7} {1.00 0.08 0.08}

\begin{document}
\title{Dynamical Casimir effect for a massless scalar field between two concentric
spherical shells with mixed boundary condictions}
\author{F. Pascoal$^{1}$, L. C. C\'{e}leri$^{2}$, S. S. Mizrahi$^{3}$, M. H. Y.
Moussa$^{1}$, C. Farina$^{4}$ \\
\\
{\small $^{1}$ Instituto de F\'{\i}sica, Universidade de S\~{a}o Paulo,}\\
{\small Caixa Postal 369, 13566-590, S\~{a}o Carlos, SP, Brazil.}\\
{\small $^{2}$Universidade Federal do ABC, Centro de Ci\^{e}ncias
Naturais e Humanas,}\\{\small R. Santa Ad\'{e}lia 166, Santo
Andr\'{e}, 09210-170,
S\~{a}o Paulo, Brazil.}\\
{\small $^{3}$ Departamento de F\'{\i}sica, CCET, Universidade Federal de S\~{a}o Carlos,}\\
{\small Via Washington Luiz km 235, 13565-905, S\~{a}o Carlos, SEP,
Brazil.}\\
{\small $^{4}$ Instituto de F\'{\i}sica, Universidade Federal do Rio
de Janeiro,}\\ {\small Caixa Postal 68528, 21945-970 Rio de Janeiro
RJ, Brazil.}} \maketitle

\begin{abstract}
We analyze the dynamical Casimir effect for a massless scalar field
confined between two concentric spherical shells which impose on the
field mixed boundary conditions. We thus complement a previous
result [Phys. Rev. A \textbf{78}, 032521 (2008)], where the same
problem was considered but in that case the field was submitted to a
Dirichlet boundary condition in both moving spherical shells. A
general expression for the average number of created particles
 is deduced for an arbitrary law of radial motion of the
spherical shells. This expression is then applied to harmonic
oscillations of the shells and the number of created particles
 is analyzed and compared with the results obtained under
Dirichlet-Dirichlet boundary conditions.

\end{abstract}

\newpage

\section{Introduction}

Since the pioneering paper published by Moore in 1970
\cite{Moore1970} and the contributions by Fulling and Davies
\cite{FullingDavies1976} and by Ford and Vilenkin
\cite{FordVilenkin1982} that appeared some years later, radiation
reaction force on moving boundaries attracted the attention of many
physicists. Due to the movement of the boundary, this topic is also
referred to as the dynamical Casimir effect (DCE), a name coined by
J. Schwinger in his attempt to explain sonoluminescence in the early
90s \cite{Schwinger1993}. For a review on this subject see the book
by K. A. Milton \cite{MiltonLivro2001} and on DCE see Refs.
\cite{DodonovRevisao2001,SpecialIssue2005}.

Though the Casimir force on a unique static plate in vacuum is zero,
the fluctuations of this force are non-vanishing \cite{Barton1991}.
Hence, if this plate starts moving with a non-zero general
acceleration, we expect that a dissipative force proportional to
these fluctuations appears
\cite{BraginskyKhalili1991,JaekelReynaud1992,MaiaNetoReynaud1993},
and arguments based on energy conservation lead directly to real
particle creation.

Though the DCE already occurs for a unique moving boundary,
oscillating cavities in parametric resonance with a particular field
mode of the corresponding static cavity may enhance significantly
the  particle creation rate
\cite{DodonovKlimov1996,LambrechtEtAl-PRL1996,JungSoh1998}. This
effect was studied by several authors considering the case of the
$1+1$ ideal cavity \cite{DodonovKlimov1996,JungSoh1998}. The $3+1$
case was also investigated, and different geometries were taken into
account, among them parallel plane plates
\cite{DodonovKlimov1996,LambrechtEtAl-PRL1996,MundurainPAMN1998},
cylindrical \cite{CrocceEtAl2005}, and spherical
\cite{Eberlein1996,SetareSaharian2001,MazzitelliMilln2006,PascoalEtAl2008}
cavities. The nonideal case was also considered in Refs.
\cite{SchallerEtAl2002,PascoalEtAlArxiv08}.

Concerning the static scenario, T. H. Boyer \cite{Boyer1974} was the
first to consider the case of mixed boundary conditions (BCs). He
demonstrated that the electromagnetic Casimir force between a
perfectly conducting plate and an infinitely permeable one is
repulsive rather than attractive. An analogous result was also
obtained in the case of a scalar field confined within two parallel
plates \cite{Hushwater1997,Cougo-PintoEtAl1999,AguiarEtAl2003} and
submitted to a Dirichlet BC at one plate and to a Neumann BC at the
other.

The measurement of a repulsive Casimir effect has been persued for
many years and has finally been achieved very recently by Munday,
Capasso and Parsegian \cite{MundayEtAl-Nature2009} in a remarkable
experiment involving three distinct media, with appropriate values
for their permitivity. Although the set up used by these authors in
their experiment on repulsive Casimir effect is quite different from
the two-plate set up made of a perfectly conducting plate and an
infinitely permeable one, we may learn many things studying the DCE
with such mixed BCs. Further, though mixed BCs are relatively common
in the study of the static Casimir effect
\cite{Hushwater1997,Cougo-PintoEtAl1999,AguiarEtAl2003,ZhaiEtAl2007,FullingEtAl2007,Teo2009},
and also in correlated topics of Cavity QED
\cite{Cougo-PintoEtAlPLB1999,Cougo-PintoEtAlJPA1999,AlvesEtAl2000,AlvesEtAl2003},
the same does not occur for the DCE. In fact, as far as we know, the
DCE in a $1+1$ dimensional resonant cavity with mixed BCs was
considered only very  recently, in Ref(s).
\cite{AlvesEtAl2006,AlvesGranhen2008}. However, mixed BCs have never
been considered in the study of DCE for different geometries as, for
instance, in concentric (and oscillating) spherical shells.

In a recent paper \cite{PascoalEtAl2008}, the DCE was examined for a
massless scalar field submitted to Dirichlet BCs at two  concentric
spherical shells, each of them posessing a time-dependent radius. A
general expression for the average number of created particles was
derived for  arbitrary laws of radial motions of the spherical
shells. Such an expression was thus applied to breathing modes of
the concentric shells: when only one of the shells oscillates and
when both shells oscillate in or out of phase. The purpose of this
paper is to complement the previous one \cite{PascoalEtAl2008} by
considering mixed BCs. We observe that the field modes associated
with mixed BCs differs from that following from
 Dirichlet-Dirichlet BCs. Considering again an oscillatory motion
of the shells, we identify all the resonances within mixed BCs and
derive the expression for the associated particle creation rate.
Then, performing a numerical analysis we compare our results with
those presented in Ref. \cite{PascoalEtAl2008}. For convenience, we
shall assume that the spherical shell which imposes a Neumann BC to
the field is at rest, while the other one, which imposes a Dirichlet
BC to the field is in arbitrary motion. However, we shall consider
two situations: in one of them, the inner shell is at rest while the
outer one is in arbitrary motion and in the other one, the reverse
occurs, namely, the outer shell is at rest while the inner one is in
arbitrary motion. Comparisons of our results with those involving
only Dirichlet-Dirichlet BCs will be presented graphically. This
paper is organized as follows: in Section 2 we briefly summarize the
main steps of the method emplyed to the case where only Dirichlet
BCs were considered; in Section 3 we apply this method to the case
of mixed BCs and obtain our general formulas; in Section 4, with the
purpose of obtaining explicit results for the average number of
created particles, we choose a particular motion for the oscillating
shell and Section 5 is left for the concluding remarks.

\section{Dirichlet-Dirichlet BCs}

In Ref \cite{PascoalEtAl2008} the DCE for a massless scalar field
confined between two cocentric moving shells was considered. This
quantum scalar field obeys the Klein-Gordon equation
$\square\phi(\mathbf{r};t)=0$. Besides, this field  and its
canonical momentum $\pi(\mathbf{r};t)=\dot{\phi}(\mathbf{r};t)$
satisfy the
equal time commutation relations
\begin{eqnarray}
\left[  \phi(\mathbf{r};t),\pi(\mathbf{r}';t)\right]   & =&
i\delta(\mathbf{r}-\mathbf{r}'),\cr\cr \left[
\phi(\mathbf{r};t),\phi(\mathbf{r}';t)\right]   & =& \left[
\pi(\mathbf{r};t),\pi(\mathbf{r}';t)\right]  =0. \label{1}%
\end{eqnarray}
The spherical symmetry of the problem leads us to the following
solution
\begin{eqnarray}
\phi(\mathbf{r};t)  &
=&\sum_{l=0}^{\infty}\sum_{m=-l}^{l}\sum_{s=1}^{\infty
}\sqrt{\frac{1}{2\omega_{ls}(t)}}F_{ls}(r;t)\left[  a_{lms}(t)~{Y}%
_{lm}(\theta,\varphi)+\text{h.c.}\right]  ,\cr\cr \pi(\mathbf{r};t)
& =&-i\sum_{l=0}^{\infty}\sum_{m=-l}^{l}\sum_{s=1}^{\infty
}\sqrt{\frac{\omega_{ls}(t)}{2}}F_{ls}(r;t)\left[  a_{lms}(t)~{Y}%
_{lm}(\theta,\varphi)-\text{h.c.}\right]  , \label{2}
\end{eqnarray}
where $\{Y_{lm}(\theta,\varphi)\}$ are the spherical harmonics and
the orthonormal radial functions satisfy the following differential
equation
\begin{equation}
\frac{1}{r^{2}}\frac{\text{d}}{\text{d}r}\left(  r^{2}\frac{\text{d}%
F_{ls}(r;t)}{\text{d}r}\right)  +\left(  \frac{\omega_{ls}^{2}(t)}{c^{2}%
}-\frac{l(l+1)}{r^{2}}\right)  F_{ls}(r;t)=0. \label{3}%
\end{equation}
\qquad Moreover, the operators $a_{lms}(t)$ and $a_{lms}^{\dag}(t)$ obey the
standard commutation relations%
\begin{eqnarray}
\left[
a_{lms}(t),a_{l^{\prime}m^{\prime}s^{\prime}}^{\dagger}(t)\right] &
=&\delta_{ll^{\prime}}\delta_{mm^{\prime}}\delta_{ss^{\prime}},\cr\cr
\left[  a_{lms}(t),a_{l^{\prime}m^{\prime}s^{\prime}}(t)\right]   &
=&\left[
a_{lms}^{\dagger}(t),a_{l^{\prime}m^{\prime}s^{\prime}}^{\dagger}(t)\right]
=0. \label{4}%
\end{eqnarray}

Through the time derivative of Eqs. (\ref{2}), together with the
Klein-Gordon equation and the canonical momentum formula, we obtain
the time evolution for the operators
\begin{eqnarray}
\dot{a}_{lms}(t)  & =& -i\omega_{ls}(t)a_{lms}(t)+\sum_{s^{\prime}}%
\mu_{l[ss^{\prime}]}(t)a_{lms^{\prime}}(t)\cr\cr & +&
\sum_{s^{\prime}}\mu_{l(ss^{\prime})}(t)a_{l(-m)s^{\prime}}^{\dag}(t),
\label{5}%
\end{eqnarray}
where the functions $\mu_{l(  ss^{\prime})  }( t)$ $=[
\mu_{lss^{\prime}}(t)+\mu_{ls^{\prime}s}(t)] /2$ and $\mu_{l[
ss^{\prime}]  }(t)=[ \mu
_{lss^{\prime}}(t)\linebreak-\mu_{ls^{\prime}s}(t)] /2$ are the
symmetric and antisymmetric parts, respectively, of the
time-dependent coefficient
\begin{eqnarray}
\mu_{lss^{\prime}}(t)  &  =&\frac{\dot{\omega}_{ls}(t)}{2\omega_{ls}(t)}%
\delta_{ss^{\prime}}\cr\cr
&  +& \left(  1-\delta_{ss^{\prime}}\right)  \sqrt{\frac{\omega_{ls}(t)}%
{\omega_{ls^{\prime}}(t)}}\int_{r_{i}(t)}^{r_{o}(t)}r^{2}F_{ls^{\prime}%
}(r;t)\dot{F}_{ls}(r;t)\operatorname*{d}r. \label{6}%
\end{eqnarray}

As demonstrated in Ref. \cite{PascoalEtAl2008}, by comparing Eq.
(\ref{5}) with the
Heisenberg equation of motion $\dot{a}_{lms}(t)=i\left[  H_{eff}%
(t),a_{lms}(t)\right]  $ and assuming the most general quadratic
form of an effective Hamiltonian, we derive
\begin{eqnarray}
H_{eff}(t) &  =& \sum_{l,m,s}\omega_{ls}(t)\left(  a_{lms}^{\dag}a_{lms}%
+\frac{1}{2}\right)  \cr\cr &
+&\frac{i}{2}\sum_{l,m,s,s^{\prime}}\mu_{lss^{\prime}}(t)\left[
\left( a_{lms^{\prime}}+a_{l\left(  -m\right)
s^{\prime}}^{\dag}\right) a_{lms}^{\dag}\right.  \cr\cr
&  -& \left.  a_{lms}\left(  a_{l\left(  -m\right)  s^{\prime}}+a_{lms^{\prime}%
}^{\dag}\right)  \right]  .\label{7}%
\end{eqnarray}
The evolution of the density operator is computed through the
relation $\dot{\rho}(t)=i\left[ H_{eff}(t),\rho(t)\right]  $, with
the aid of an iterative procedure up to second order approximation
in the velocity of the cavity boundaries, \emph{i.e.},
$\dot{r}_{i}(t)$, $\dot{r}_{0}(t)\ll c$. The derivation of the
average number of particles created in a particular mode --- labeled
by the quantum numbers ($l,m,s$) --- is thus given by
$\mathcal{N}_{lms}(t)=Tr\left[
\rho(t)a_{lms}^{\dag}(0)a_{lms}(0)\right]  $, and for an initial
vacuum state $\rho(0)=\left\vert \left\{  0\right\}  \right\rangle
\left\vert \left\{ 0\right\}  \right\rangle $ it follows that
\begin{equation}
\mathcal{N}_{lms}(t)=\sum_{s^{\prime}}\left\vert \int_{0}^{t}\text{d}t_{1}%
\mu_{l\left(  s^{\prime}s\right)  }(t_{1})\exp\left\{  i\left[  \Omega
_{ls^{\prime}}(t_{1})+\Omega_{ls}(t_{1})\right]  \right\}  \right\vert
^{2},\label{8}%
\end{equation}
with $\Omega_{ls}(t)=\int_{0}^{t}dt_{1}\ \omega_{ls}(t_{1})$.

The number of created particles $\mathcal{N}_{lms}(t)$ depends on
the radial function $F_{ls}(r;t)$ through $\mu_{l\left(
s^{\prime}s\right)  }(t)$. The solution of Eq. (\ref{3}) is given by
a linear combination of spherical Bessel functions of the first
($j_{l}$) and second ($n_{l}$) kind, such that the
Dirichlet BC applied to the inner shell leads to the relation
\begin{equation}
F_{ls}(r;t)=N_{ls}\bigl[  j_{l}\bigl(  \omega_{ls}(t)r\bigr)
n_{l}\bigl( \omega_{ls}(t)r_{i}(t)\bigr)  -j_{l}\bigl(
\omega_{ls}(t)r_{i}(t)\bigr)
n_{l}\bigl(  \omega_{ls}(t)r\bigr)  \bigr]  ,\label{9}%
\end{equation}
whereas that on the outer shell results in the transcendental equation%
\begin{equation}
j_{l}\bigl(  \omega_{ls}(t)r_{o}(t)\bigr)  n_{l}\bigl(  \omega_{ls}%
(t)r_{i}(t)\bigr)  -j_{l}\bigl(  \omega_{ls}(t)r_{i}(t)\bigr)
n_{l}\bigl(
\omega_{ls}(t)r_{o}(t)\bigr)  =0\text{.}\label{10}%
\end{equation}
In Fig. 1 we present a map of the solutions of Eq. (\ref{10}) for
some values of the numbers $l$ and $s$. As it was noted in
\cite{PascoalEtAl2008}, for the case $l=0$, the frequencies are
equally spaced. This fact does not occur for the case $l\neq0$.
However, when both radii of the shells are much larger than the
separation between them, \textit{i.e.}, $r_{i}\left( t\right)
>>r_{o}\left(  t\right) -r_{i}$ $\left( t\right)  $, the solutions
for all values of $l$ approach the solution for the onedimensional
case, so that $\omega_{ls}\rightarrow s\pi/\left( r_{o}\left(
t\right) -r_{i}\left(  t\right)  \right)  $.

\begin{figure}[!h]
\begin{center}
\newpsobject{showgrid}{psgrid}{subgriddiv=0.5,griddots=10,gridlabels=6pt}

\begin{pspicture}(-1,0.2)(3,4.5)
\psset{unit=1.15}
%
\psline[linecolor=c0](0,1)(3,4)%
\pscurve[linecolor=c1](0,1.4303)(0.1,1.43356) (0.2,1.45278) (0.4,1.54946) (0.6,1.69621) (1,2.04904) (1.5,2.52667) (2,3.01678) (2.991,4)%
\pscurve[linecolor=c2](0,1.83457)(0.1,1.83459) (0.2,1.83526) (0.4,1.85073) (0.6,1.91003) (1, 2.15883) (1.5, 2.58441) (2, 3.05225) (2.9738,4)%
\pscurve[linecolor=c3](0,2.22433)(0.1,2.22433) (0.2,2.22434) (0.4,2.22521) (0.6,2.23499) (1, 2.35059) (1.5, 2.68295) (1.8132801537026406, 2.9449) (2.9449,4)%
\pscurve[linecolor=c4](0,2.60459)(0.1,2.60459) (0.2,2.60459) (0.4,2.60461) (0.6,2.60533) (1, 2.63536) (1.5, 2.83813) (2, 3.19951) (2.9012,4)%
\pscurve[linecolor=c5](0,2.97805)(0.1,2.97805) (0.2,2.97805) (0.4,2.97805) (0.6,2.97808) (1, 2.98234) (1.5, 3.06693) (2, 3.33035) (2.83535,4)%
\pscurve[linecolor=c6](0,3.34634)(0.1,3.34634) (0.2,3.34634) (0.4,3.34634) (0.6,3.34634) (1, 3.34671) (1.5, 3.36892) (2, 3.51818) (2.5,3.82504) (2.73035,4)%
\pscurve[linecolor=c7](0,3.71055)(0.1,3.71055) (0.2,3.71055) (0.4,3.71055) (0.6,3.71055) (1,3.71057) (1.5, 3.71425) (2, 3.77358) (2.52949,4)%
\psline[linecolor=c0,linestyle=dashed](0,2)(2,4)%
\pscurve[linecolor=c1,linestyle=dashed](0,2.45902)(0.1,2.46218) (0.2,2.48081) (0.4,2.5748) (0.6,2.71815) (1,3.06516) (1.5,3.538) (1.97,4)%
\pscurve[linecolor=c2,linestyle=dashed](0,2.89503) (0.1,2.89505) (0.2,2.89568) (0.4,2.91031) (0.6,2.96653) (1, 3.20468) (1.5, 3.61797) (1.91768, 4)%
\pscurve[linecolor=c3,linestyle=dashed](0,3.31587)(0.1,3.31587) (0.2,3.31588) (0.4,3.31669) (0.6,3.32575) (1,3.43349) (1.5,3.74823) (1.8132801537026406,4.)%
\pscurve[linecolor=c4,linestyle=dashed](0,3.72579)(0.1,3.72579) (0.2,3.72579) (0.4,3.72581) (0.6,3.72646) (1, 3.75389)(1.59989,4)%
\psline[linecolor=c0,linewidth=.06,linestyle=dotted,dotsep=1pt](0,3)(1,4)%
\pscurve[linecolor=c1,linewidth=.06,linestyle=dotted,dotsep=1pt](0,3.47089)(0.1,3.47402) (0.2,3.4925)(0.4,3.58577) (0.6,3.72815) (0.915,4)%
\pscurve[linecolor=c2,linewidth=.06,linestyle=dotted,dotsep=1pt](0,3.92251) (0.1,3.92254) (0.2,3.92315) (0.4,3.93754) (0.6,3.99285) (0.617497,4)%
\pspolygon[linewidth=0.04](0,1)(3,1)(3,4)(0,4)%
\psline[linewidth=0.04] (0,1)(-0.2,1) \rput(-0.5,1){$1$}%
\psline[linewidth=0.04] (0,1.5)(-0.15,1.5)%
\psline[linewidth=0.04] (0,2)(-0.2,2)\rput(-0.5,2){$2$}%
\psline[linewidth=0.04] (0,2.5)(-0.15,2.5)%
\psline[linewidth=0.04] (0,3)(-0.2,3)\rput(-0.5,3){$3$}%
\psline[linewidth=0.04] (0,3.5)(-0.15,3.5)%
\psline[linewidth=0.04] (0,4)(-0.2,4)\rput(-0.5,4){$4$}%
\rput{90}(-1,2.5){$\omega_{ls} r_o/\pi$}%
\psline[linewidth=0.04] (0,1)(0,0.8) \rput(0,0.5){$0$}%
\psline[linewidth=0.04] (0.5,1)(0.5,0.85)%
\psline[linewidth=0.04] (1,1)(1,0.8) \rput(1,0.5){$1$}%
\psline[linewidth=0.04] (1.5,1)(1.5,0.85)%
\psline[linewidth=0.04] (2,1)(2,0.8) \rput(2,0.5){$2$}%
\psline[linewidth=0.04] (2.5,1)(2.5,0.85)%
\psline[linewidth=0.04] (3,1)(3,0.8) \rput(3,0.5){$3$}%
\rput(1.50,0.25){$\omega_{ls} r_i/\pi$}%
\end{pspicture}
\end{center}
\vskip -0.6 cm
 \caption{Map of the solutions of the transcendental
equation (\ref{10}). The colors correspond to different values of
the number $l$. The solid, dashed, and dotted lines correspond to
$s=1$, $s=2$, and $s=3$, respectively.} \label{Fa1}
\end{figure}
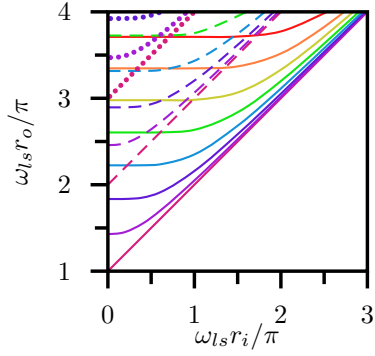

\section{Mixed boundary conditions}

It is important to emphasize that the expression for the average
number of created particles, derived in Eq (\ref{8}), does not
depend on the character of the BCs. Thus it can be applied even for
mixed BCs as in the present work, where we assume that the massaless
scalar field satisfies the Neumann BC at a fixed spherical shell and
a Dirichlet BC at a second concentric spherical shell whose radius
has an arbitrary time dependence,
\begin{equation}
  \partial_{r}\phi\left(  \mathbf{r},t\right)  \vert
_{r=r_{\alpha}}=   0\;\;\;\;\;
 \mbox{and}\;\;\;\;\;      \left.  \phi\left(
\mathbf{r},t\right) \right\vert
_{r=r_{\beta}(t)}=0, \label{11}%
\end{equation}
where the index $\alpha$ ($\beta$) is related to the static (moving)
shell.\textbf{ }However, different expressions come up for
$F_{ls}(r;t)$ and $\omega_{ls}(t)$, as compared to those in Ref.
(\ref{16}).

As already noted in the previous section, the general solutions to
Eq. (\ref{3}) are linear combinations of spherical Bessel functions,
but the mixed BC leads to a different solution. The assumption of
Neumann BC for the field at the static shell leads to the following
expression for the radial functions
\begin{equation}
F_{ls}(r;t)=N_{ls}\left(  j_{l}\left(  \omega_{ls}(t)r\right)  \ \frac
{\partial}{\partial r_{\alpha}}n_{l}\left(  \omega_{ls}(t)r_{\alpha}\right)
-n_{l}\left(  \omega_{ls}(t)r\right)  \frac{\partial}{\partial r_{\alpha}%
}j_{l}\left(  \omega_{ls}(t)r_{\alpha}\right)  \right)  , \label{12}%
\end{equation}
and the subsequent assumption of Dirichlet BC on the field at the
moving shell leads to a frequency discretization
\begin{equation}
\frac{\partial}{\partial r_{\alpha}}j_{l}\bigl(
\omega_{ls}(t)r_{\alpha }\bigr)  \ n_{l}\bigl(
\omega_{ls}(t)r_{\beta}(t)\bigr)  =j_{l}\bigl(
\omega_{ls}(t)r_{\beta}(t)\bigr)  \ \frac{\partial}{\partial r_{\alpha}}%
n_{l}\bigl(  \omega_{ls}(t)r_{\alpha}\bigr)  . \label{13}%
\end{equation}

In Figs. 2 and 3 we show the maps of the numerical solutions of the
transcendental equation  Eq. (\ref{13}) for some values of $l$ and
$s$. As we can see, the map of $\omega_{ls}(t)$ is very sensitive to
the BCs. For Dirichlet-Dirichlet BCs (Fig. 1) the frequencies
$\omega_{0s}(t)$ are equally spaced, a situation that does not occur
for mixed BCs. We also note that the map of $\omega_{ls}(t)$ turns
out to be entirely different when considering the Dirichlet BC in
the outer shell and Neumann BC in the inner one (Fig. 2) or
oppositely, with Neumann BC in the outer shell and Dirichlet BC in
the inner one (Fig. 3).

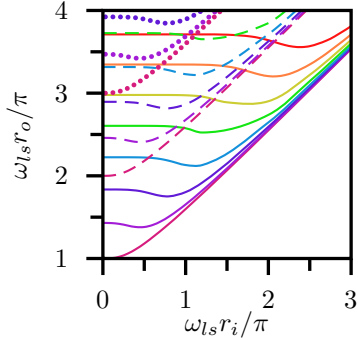
\begin{figure}[!h]
\begin{center}
\newpsobject{showgrid}{psgrid}{subgriddiv=0.5,griddots=10,gridlabels=6pt}

\begin{pspicture}(-1,0)(3,4.5)
\psset{unit=1.10}
%
\pscurve[linecolor=c0](0,1)(0.1,1.00311)(0.2,1.02143)(0.3,1.05942)(0.4,1.11395)(0.5,1.18045)(0.75,1.37776)(1,1.59809) (1.5,2.06656) (2,2.55024) (2.5,3.04031) (3,3.53365)%
\pscurve[linecolor=c1](0,1.4303)(0.1,1.42862) (0.2,1.41831) (0.4,1.38069) (0.5,1.38087) (0.6,1.40468) (0.7,1.44886) (0.8, 1.50792) (1, 1.65395) (1.37, 1.97193) (2, 2.56366) (2.44,2.99022) (3,3.5395)%
\pscurve[linecolor=c2](0,1.83457)(0.1,1.83455) (0.2,1.8341) (0.4,1.82268) (0.5,1.80496) (0.6,1.78) (0.7,1.75805) (0.8, 1.75192) (1, 1.80326) (1.2, 1.91766) (2, 2.59321) (2.4,2.97098) (3,3.55174)%
\pscurve[linecolor=c3](0,2.22433)(0.1,2.22433) (0.2,2.22432) (0.4,2.22364) (0.5,2.22144) (0.6,2.21548) (0.7,2.20286) (0.8,2.18173) (1, 2.13053) (1.2,2.13018) (2,2.64584) (2.5,3.09727) (3,3.57161)%
\pscurve[linecolor=c4](0,2.60459)(0.1,2.60459) (0.2,2.60459) (0.4,2.60457) (0.5,2.60446) (0.6,2.60398) (0.7,2.60246) (0.8,2.59859) (1, 2.57524) (1.2,2.52396) (2,2.73707) (2.5,3.14515) (3,3.60134)%
\pscurve[linecolor=c5](0,2.97805)(0.1,2.97805) (0.2,2.97805) (0.4,2.97805) (0.5,2.97804) (0.6,2.97802) (0.7,2.97792) (0.8,2.97758) (0.9,2.97659) (1, 2.97415) (1.1, 2.96891) (1.2,2.95896) (1.5,2.88915) (2,2.9021) (2.5,3.22103) (3,3.64495)%
\pscurve[linecolor=c6](0,3.34634) (0.2,3.34634) (0.4,3.34634) (0.6,3.34634)(0.8,3.34632)(1, 3.34601) (1.2,3.34387) (1.4,3.33427) (1.6,3.30532) (1.8,3.25103) (1.9,3.22344) (2,3.20572) (2.1,3.20356) (2.2,3.21851) (2.3,3.24913) (2.5,3.34688) (2.7,3.47773) (3,3.70982)%
\pscurve[linecolor=c7](0,3.71055) (0.3,3.71055) (0.6,3.71055) (0.9,3.71054) (1.2,3.71033) (1.5,3.707) (1.8,3.68373) (1.9,3.66521) (2,3.63992) (2.1,3.6102) (2.2,3.58183) (2.3,3.56225) (2.4,3.55708) (2.5,3.5683) (2.7,3.63487) (2.8,3.68545) (3,3.81038)%
\pscurve[linecolor=c0,linestyle=dashed](0,2)(0.1,2.00311)(0.2,2.02143)(0.3,2.05942)(0.4,2.11395)(0.5,2.18045)(0.75,2.37776)(1,2.59809) (1.5,3.06656) (2,3.55024) (2.459,4)%
\pscurve[linecolor=c1,linestyle=dashed](0, 2.45902)(0.1, 2.4574) (0.2,2.44741) (0.4,2.41103) (0.5, 2.41121) (0.6,2.43422) (0.7, 2.47701) (0.8, 2.53437) (1, 2.67679) (1.37, 2.9891) (2, 3.57471) (2.44,3.99869)%
\pscurve[linecolor=c2,linestyle=dashed](0,2.89503)(0.1, 2.89502) (0.2,2.89459) (0.4,2.88381) (0.5, 2.8671) (0.6,2.84361) (0.7, 2.823) (0.8, 2.81725) (1, 2.8655) (1.2,  2.97378) (2, 3.62658) (2.4,3.99723)%
\pscurve[linecolor=c3,linestyle=dashed] (0,3.31587)(0.1, 3.31587) (0.2,3.31587) (0.4,3.31524) (0.5,3.3132) (0.6,3.30768) (0.7,3.29601) (0.8,3.2765) (1,3.22941) (1.2, 3.22909) (2, 3.71278) (2.33525,4)%
\pscurve[linecolor=c4,linestyle=dashed] (0,3.72579)(0.1,3.72579) (0.2,3.72579) (0.4,3.72577) (0.5,3.72567) (0.6,3.72523) (0.7,3.72384) (0.8,3.72032) (1, 3.69907) (1.2,3.65259) (2,3.84739) (2.21292 ,4)%
\pscurve[linecolor=c0,linewidth=.06,linestyle=dotted,dotsep=1pt](0,3)(0.1,3.00311)(0.2,3.02143)(0.3,3.05942)(0.4,3.11395)(0.5,3.18045)(0.75,3.37776)(1,3.59809) (1.43,4.)%
\pscurve[linecolor=c1,linewidth=.06,linestyle=dotted,dotsep=1pt](0,3.47089)(0.1,3.46928) (0.2,3.45937) (0.4,3.4233) (0.5,3.42348) (0.6,3.4463) (0.7,3.48873) (0.8, 3.54564) (1, 3.68706) (1.37, 3.99758)%
\pscurve[linecolor=c2,linewidth=.06,linestyle=dotted,dotsep=1pt](0,3.92251)(0.1, 3.9225) (0.2,3.92208) (0.4,3.91148) (0.5,3.89506) (0.6,3.87198) (0.7,3.85173) (0.8, 3.84608) (1, 3.89348) (1.2, 3.99998)%
\pspolygon[linewidth=0.04](0,1)(3,1)(3,4)(0,4)%
\psline[linewidth=0.04] (0,1)(-0.2,1) \rput(-0.5,1){$1$}%
\psline[linewidth=0.04] (0,1.5)(-0.15,1.5)%
\psline[linewidth=0.04] (0,2)(-0.2,2)\rput(-0.5,2){$2$}%
\psline[linewidth=0.04] (0,2.5)(-0.15,2.5)%
\psline[linewidth=0.04] (0,3)(-0.2,3)\rput(-0.5,3){$3$}%
\psline[linewidth=0.04] (0,3.5)(-0.15,3.5)%
\psline[linewidth=0.04] (0,4)(-0.2,4)\rput(-0.5,4){$4$}%
\rput{90}(-1,2.5){$\omega_{ls} r_o/\pi$}%
\psline[linewidth=0.04] (0,1)(0,0.8) \rput(0,0.5){$0$}%
\psline[linewidth=0.04] (0.5,1)(0.5,0.85)%
\psline[linewidth=0.04] (1,1)(1,0.8) \rput(1,0.5){$1$}%
\psline[linewidth=0.04] (1.5,1)(1.5,0.85)%
\psline[linewidth=0.04] (2,1)(2,0.8) \rput(2,0.5){$2$}%
\psline[linewidth=0.04] (2.5,1)(2.5,0.85)%
\psline[linewidth=0.04] (3,1)(3,0.8) \rput(3,0.5){$3$}%
\rput(1.5,0.2){$\omega_{ls} r_i/\pi$}%
\end{pspicture}
\end{center}
\vskip -0.7cm
 \caption{Map of the solutions of the transcendental
equation (\ref{13}) with $r_{\alpha}<r_{\beta}$.  The colors
correspond to different values of the number $l$. The solid, dashed,
and dotted lines correspond to $s=1$, $s=2$, and $s=3$,
respectively.. }
\end{figure}

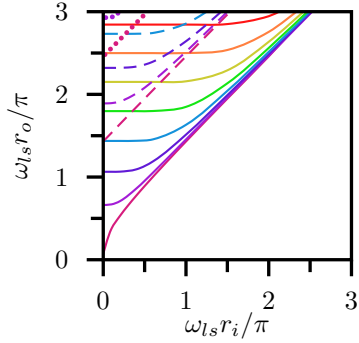
\begin{figure}[!h]
\begin{center}
\newpsobject{showgrid}{psgrid}{subgriddiv=0.5,griddots=10,gridlabels=6pt}
\psset{unit=1.1}
\begin{pspicture}(-1,-1)(3,2.5)
%
\pscurve[linecolor=c0](0,0.0678391)(0.1,0.376665) (0.2,0.52706) (0.3,0.656247)(0.4,0.77611) (0.5,0.890754) (0.75,1.16511) (1,1.4303) (1.5,1.94845) (2,2.45902) (2.5,3)%
\pscurve[linecolor=c1](0,0.662586)(0.1,0.669501) (0.2,0.707391) (0.4,0.861276) (0.6,1.05239) (1,1.45461) (1.5,1.9615) (2,2.46721) (2.525,3)%
\pscurve[linecolor=c2](0,1.06382)(0.1,1.06386) (0.2,1.06511) (0.4,1.09307) (0.6,1.18863) (1,1.51316) (1.5,1.99051) (2,2.48475) (2.51658,3)%
\pscurve[linecolor=c3](0,1.43688)(0.1,1.43688) (0.2,1.4369) (0.4,1.43846) (0.6,1.45558) (1,1.63182) (1.5,2.04285) (2,2.51437) (2.49728,3)%
\pscurve[linecolor=c4](0,1.7974)(0.1,1.7974) (0.2,1.7974) (0.4,1.79744) (0.6,1.79869) (1,1.84903) (1.5,2.13393) (2,2.5614) (2.46703,3)%
\pscurve[linecolor=c5](0,2.15065)(0.1,2.15065) (0.2,2.15065) (0.4,2.15065) (0.6,2.1507) (1,2.15801) (1.5,2.29079) (2,2.63587) (2.41879,3)%
\pscurve[linecolor=c6](0,2.49908)(0.1,2.49908) (0.2,2.49908) (0.4,2.49908) (0.6,2.49908) (1,2.4997) (1.5,2.53677) (2,2.75568) (2.33379,3)%
\pscurve[linecolor=c7](0,2.84405)(0.1,2.84405) (0.2,2.84405) (0.4,2.84405) (0.6,2.84405) (1,2.84408) (1.5,2.85033) (2,2.94523) (2.12463,3)%
\pscurve[linecolor=c0,linestyle=dashed](0,1.43135) (0.1,1.53491) (0.2,1.63894) (0.3,1.74249) (0.4,1.84564) (0.5,1.94845) (0.75,2.20435) (1,2.45902) (1.25,2.71282) (1.5,2.96597)%
\pscurve[linecolor=c1,linestyle=dashed](0,1.89088)(0.1,1.89418) (0.2,1.91366) (0.4,2.01159) (0.6,2.16013) (1,2.51673) (1.5,3)%
\pscurve[linecolor=c2,linestyle=dashed](0,2.32046)(0.1,2.32048) (0.2, 2.32114) (0.4, 2.33649) (0.6,2.39534) (1,2.64289) (1.42484,3)%
\pscurve[linecolor=c3,linestyle=dashed] (0, 2.73229)(0.1, 2.73229) (0.2, 2.7323) (0.4, 2.73316) (0.6, 2.7427) (1, 2.85585)(1.248764177214013,3)%
\pscurve[linecolor=c0,linewidth=.06,linestyle=dotted,dotsep=1pt](0,2.46004) (0.1,2.56063) (0.2,2.66212) (0.3,2.7635) (0.4,2.86478) (0.5,2.96597)%
\pscurve[linecolor=c1,linewidth=.06,linestyle=dotted,dotsep=1pt](0,2.93031)(0.1,2.93349) (0.2,3)%
\pspolygon[linewidth=0.04](0,0)(3,0)(3,3)(0,3)%
\psline[linewidth=0.04] (0,0)(-0.2,0)\rput(-0.5,0){$0$}%
\psline[linewidth=0.04] (0,.5)(-0.15,0.5)%
\psline[linewidth=0.04] (0,1)(-0.2,1) \rput(-0.5,1){$1$}%
\psline[linewidth=0.04] (0,1.5)(-0.15,1.5)%
\psline[linewidth=0.04] (0,2)(-0.2,2)\rput(-0.5,2){$2$}%
\psline[linewidth=0.04] (0,2.5)(-0.15,2.5)%
\psline[linewidth=0.04] (0,3)(-0.2,3)\rput(-0.5,3){$3$}%
\rput{90}(-1,1.5){$\omega_{ls} r_o/\pi$}%
\psline[linewidth=0.04] (0,0)(0,-0.2) \rput(0,-0.5){$0$}%
\psline[linewidth=0.04] (0.5,0)(0.5,-0.15)%
\psline[linewidth=0.04] (1,0)(1,-0.2) \rput(1,-0.5){$1$}%
\psline[linewidth=0.04] (1.5,0)(1.5,-0.15)%
\psline[linewidth=0.04] (2,0)(2,-0.2) \rput(2,-0.5){$2$}%
\psline[linewidth=0.04] (2.5,0)(2.5,-0.15)%
\psline[linewidth=0.04] (3,0)(3,-0.2) \rput(3,-0.5){$3$}%
\rput(1.5,-0.8){$\omega_{ls} r_i/\pi$}%
\end{pspicture}
\end{center}
\vskip -0.6cm
 \caption{Map of the solutions of the transcendental
equation (\ref{13}) with $r_{\alpha}>r_{\beta}$.  The colors
correspond to different values of the number $l$. The solid, dashed,
and dotted lines correspond to $s=1$, $s=2$, and $s=3$,
respectively.}
\end{figure}

However, there are also some similarities between the results
derived from mixed BCs and those for Dirichlet-Dirichlet BCs, in
Fig. 1; it can be directly verified that the solutions for
$\omega_{ls}(t)$, coming from Eq. (13), approach that for the
onedimensional case ($\omega_{ls}\rightarrow\left( s-1/2\right)
\pi/\left\vert r_{\alpha }-r_{\beta}\left(  t\right) \right\vert $)
when both radii of the shells are much larger than the separation
between them.

A comment is in order here: we note that the BCs must be imposed in
the instantaneously co-moving Lorentz frame, where the boundaries
are momentarily at rest. If the Neumann BC was to be imposed on the
moving boundary, we should have used the appropriate Lorentz
transformation to write the fields in the
inertial frame of the laboratory as follow\textbf{s }
\begin{equation}
\left.  \partial_{r^{\prime}}\phi\left( \mathbf{r}^{\prime},t\right)
\right\vert
_{r^{\prime}=r_{\beta}^{\prime}(t)}
 \;\;\;\Longrightarrow\;\;\;
  \left.\left\{
\partial_{r}+\dot{r}_{\beta}(t)\partial_{t}\right\}  \phi\left(
\mathbf{r},t\right)  \right\vert _{r=r_{\beta}(t)}=0.\label{13b}%
\end{equation}
In that case, the time derivative in Eq. (\ref{13b}) invalidates the
expansion used in Eq. (\ref{2}), and as a consequence, also in Eq.
(\ref{8}). This fact demands a different formal development for the
computation of the required particle creation rate. For that reason,
in the present work we treat only the case where the Neumann BC is
imposed on a spherical shell at rest, leaving aside the breathing
modes analyzed in Ref. \cite{PascoalEtAl2008}, when both shells
oscillate in or out of phase.

\section{Numerical estimatives}

In this section, in order to obtain explicit results, we will
consider an specific motion for the spherical shell that imposes on
the field the Dirichlet BC. A typical situation consists of an
oscillation that starts at some instant, has a sinusoidal behaviour
with an angular frequency $\varpi $ and a small amplitude and then
stops at some later instant. We, then, assume that the radius of the
moving shell has the following law of motion
\begin{equation}
r_{\beta}(t)=r_{\beta}\left(  1+\epsilon\sin\left(  \varpi t\right)  \right)
,\label{14}%
\end{equation}
with $\epsilon \ll 1$. In the following we also assume that the
cavity mirror oscillates only during a finite time interval $T$,
then stopping suddenly its motion.

Substituting Eqs. (\ref{6}) and (\ref{14}) into Eq. (\ref{8}) and
making a power series expansion with respect to the small parameter
$\epsilon$, we obtain
\begin{equation}
\mathcal{N}_{lms}=\left(  \frac{\epsilon\varpi T}{2}\right)  ^{2}%
\sum_{s^{\prime}}\left\vert C_{l\left( ss^{\prime}\right)}
 \,f_{lss^{\prime}}\left(  \varpi;T\right)  \right\vert ^{2},
  \label{15}
\end{equation}
where, after defining
$\omega_{lss^{\prime}}\equiv\omega_{ls}(0)+\omega
_{ls^{\prime}}(0)$, the coefficient $C_{lss^{\prime}}$ and function
$f_{lss^{\prime}}\left(  \varpi;T\right)  $ are given by
\begin{equation}
f_{lss^{\prime}}\left(  \varpi;T\right)  =\frac{\exp\left[  i\left(
\varpi-\omega_{lss^{\prime}}\right)  T\right]  -1}{i\left(  \varpi
-\omega_{lss^{\prime}}\right)  T}-\frac{\exp\left[  -i\left(  \varpi
+\omega_{lss^{\prime}}\right)  T\right]  -1}{i\left(  \varpi+\omega
_{lss^{\prime}}\right)  T}, \label{16}%
\end{equation}
and
\begin{eqnarray}
C_{lss^{\prime}}  &  =&r_{\beta}\delta_{ss^{\prime}}\frac{1}{2\omega_{ls}%
(0)}\frac{\partial\omega_{ls}(0)}{\partial r_{\beta}}\cr\cr & -&
r_{\beta}\left(  1-\delta_{ss^{\prime}}\right)  \sqrt{\frac{\omega
_{ls}(0)}{\omega_{ls^{\prime}}(0)}}\int_{r_{i}}^{r_{o}}dr~r^{2}F_{ls}%
(r;0)\frac{dF_{ls^{\prime}}(r;0)}{dr_{\beta}}. \label{17}%
\end{eqnarray}

Note that $f_{lss^{\prime}}\left(  \varpi;T\right)  $ is an
oscillating function of $T$, except when the mirror oscillating
frequency satisfies the resonance condition, namely,
$\varpi=\omega_{lss^{\prime}}$. In this case $f_{lss^{\prime}}\left(
\omega_{lss^{\prime}};T\right)  =1$, and
the number of created particles turns to be the following quadratic function of $T$,
\begin{equation}
\lim_{\omega\rightarrow\omega_{lss^{\prime}}}\mathcal{N}_{lms}=\left(
\frac{\epsilon\omega_{lss^{\prime}}T}{2}\right)  ^{2}\left\vert
C_{l\left( ss^{\prime}\right)  }\right\vert ^{2}.
 \label{18}
\end{equation}
This result is  valid only under the short-time approximation
$\epsilon \omega_{lss^{\prime}}T<<1$, since we have disregard terms
proportional to $\left(  \epsilon\omega_{lss^{\prime}}T\right) ^{n}$
with $n\geq3$. Moreover, Eqs. (\ref{15}) to (\ref{18}) are valid
either for Dirichlet-Dirichlet BCs or mixed BCs since they were
derived using the equation of motion (\ref{14}) and equations
(\ref{1}) - (\ref{8}) which are independent from the BCs.

To study the behavior of our result in Eq. (\ref{18}), we plot in
Fig. 4 the expression $\mathcal{N}_{lms}/(\epsilon\varpi T)^{2}$ as
a function of $\varpi\pi/(r_{o}-r_{i})$ for some values of $l$ and
$s$, under resonance conditions $\varpi=\omega_{lss^{\prime}}$,
setting $r_{o}=2r_{i}$. Both letters in the legend indicate the BCs
on the inner and the outer shells, respectively: for example, $D$
($\tilde{D}$) means Dirichlet BC on a static (moving) shell, whereas
$N$ means Neumann BC on a static shell. As we can see, both the
intensity and position of the resonances change in a non-trivial way
with the BC. The case  of a moving  outer shell with the field
satisfying Dirichlet-Dirichlet BCs exhibits higher resonance
intensities, while the case of a moving inner shell with the field
submitted to mixed BCs leads to lower resonance frequencys.

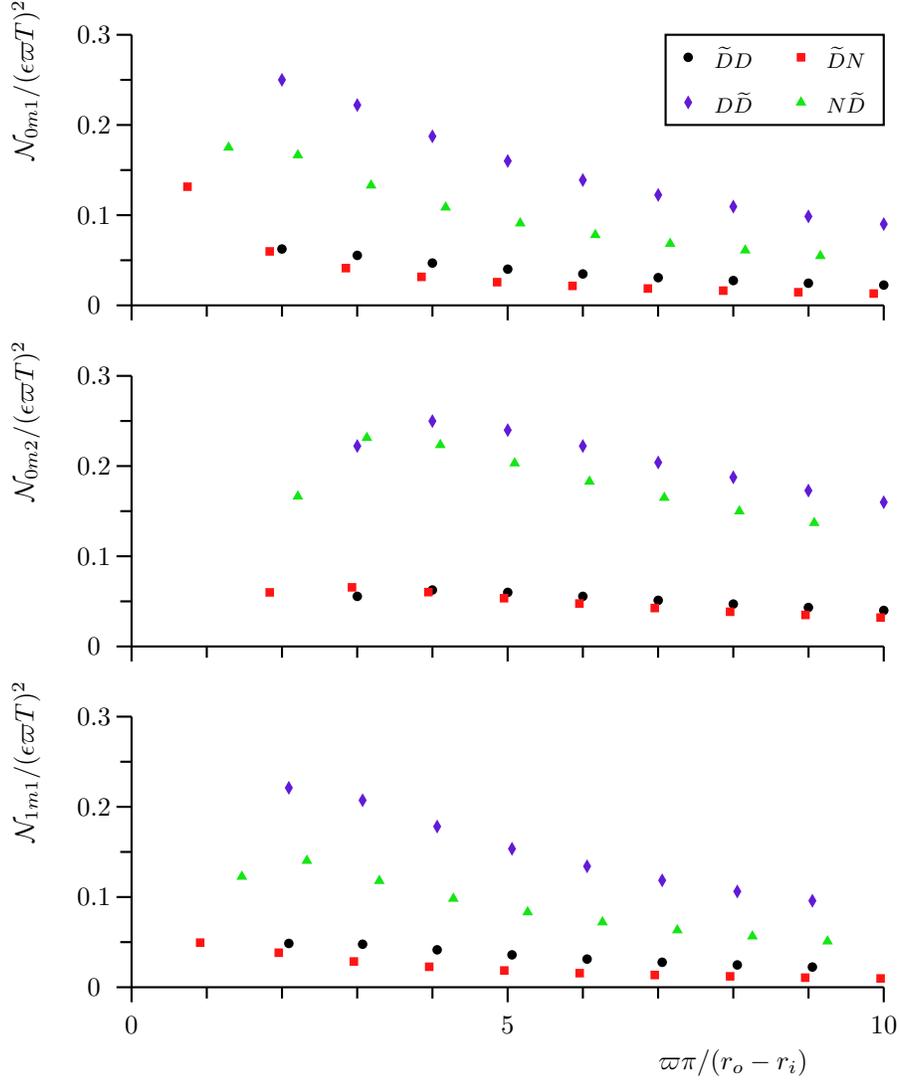
\begin{figure}[!h]
\begin{center}
\newpsobject{showgrid}{psgrid}{subgriddiv=0.5,griddots=10,gridlabels=6pt}
\psset{xunit=1} \psset{yunit=3}
\begin{pspicture}(-1,-0.3)(10,1.2)
\psline(-0.2,0)(10,0) \rput(-0.5,0){$0$}%
\psline(-0.15,0.2)(0,0.2)%
\psline(-0.2,0.4)(0,0.4) \rput(-0.5,0.4){$0.1$}%
\psline(-0.15,0.6)(0,0.6)%
\psline(-0.2,0.8)(0,0.8) \rput(-0.5,0.8){$0.2$}%
\psline(-0.15,1)(0,1)%
\psline(-0.2,1.2)(0,1.2) \rput(-0.5,1.2){$0.3$}%
\rput{90}(-1.4,1.0){${\cal N}_{0m1}/(\epsilon \varpi T)^2$}
\psline(0,-0.066666666)(0,1.2)%
\psline(1,-0.05)(1,0)%
\psline(2,-0.05)(2,0)%
\psline(3,-0.05)(3,0)%
\psline(4,-0.05)(4,0)%
\psline(5,-0.066666666)(5,0)%
\psline(6,-0.05)(6,0)%
\psline(7,-0.05)(7,0)%
\psline(8,-0.05)(8,0)%
\psline(9,-0.05)(9,0)%
\psline(10,-0.06666666)(10,0)%
%
\psdots (2., 0.25) (3., 0.222222) (4., 0.1875) (5., 0.16) (6., 0.138889) (7., 0.122449) (8., 0.109375) (9., 0.0987654) (10., 0.09)%
\psdots[linecolor=c2,dotstyle=diamond*] (2., 1.) (3., 0.888889) (4., 0.75) (5., 0.64) (6., 0.555556) (7., 0.489796) (8., 0.4375) (9., 0.395062) (10., 0.36)%
\psdots[linecolor=c7,dotstyle=square*](0.742019, 0.526384) (1.83658, 0.239496) (2.85061, 0.165635) (3.85648, 0.126694) (4.85973, 0.102579) (5.86179, 0.0861728) (6.86321, 0.0742893) (7.86425, 0.065285) (8.86505, 0.0582269) (9.86567, 0.0525456)%
\psdots[linecolor=c4,dotstyle=triangle*] (1.29155, 0.699717) (2.20969, 0.665598) (3.18546, 0.532564) (4.17441, 0.434037) (5.16814, 0.364052) (6.16411, 0.312774) (7.16131, 0.273863) (8.15925, 0.243425) (9.15767, 0.219004)%
\pspolygon(10,1.2)(10,0.8)(7.1,0.8)(7.1,1.2)%
\psdot (7.4,1.1)%
\psdot[linecolor=c2,dotstyle=diamond*] (7.4,0.9)%
\psdot[linecolor=c7,dotstyle=square*] (8.9,1.1)%
\psdot[linecolor=c4,dotstyle=triangle*] (8.9,0.9)%
\rput (8.0,1.1) {\footnotesize $\widetilde{D} D$}%
\rput (8.0,0.9) {\footnotesize $D \widetilde{D}$}%
\rput (9.5,1.1) {\footnotesize $\widetilde{D} N$}%
\rput (9.5,0.9) {\footnotesize $N \widetilde{D}$}%
\end{pspicture}
\begin{pspicture}(-1,-0.3)(10,1.2)
\psline(-0.2,0)(10,0) \rput(-0.5,0){$0$}%
\psline(-0.15,0.2)(0,0.2)%
\psline(-0.2,0.4)(0,0.4) \rput(-0.5,0.4){$0.1$}%
\psline(-0.15,0.6)(0,0.6)%
\psline(-0.2,0.8)(0,0.8) \rput(-0.5,0.8){$0.2$}%
\psline(-0.15,1)(0,1)%
\psline(-0.2,1.2)(0,1.2) \rput(-0.5,1.2){$0.3$}%
\rput{90}(-1.4,1.0){${\cal N}_{0m2}/(\epsilon \varpi T)^2$}
\psline(0,-0.066666666)(0,1.2)%
\psline(1,-0.05)(1,0)%
\psline(2,-0.05)(2,0)%
\psline(3,-0.05)(3,0)%
\psline(4,-0.05)(4,0)%
\psline(5,-0.066666666)(5,0)%
\psline(6,-0.05)(6,0)%
\psline(7,-0.05)(7,0)%
\psline(8,-0.05)(8,0)%
\psline(9,-0.05)(9,0)%
\psline(10,-0.06666666)(10,0)%
\psdots (3., 0.222222) (4., 0.25) (5., 0.24) (6., 0.222222) (7., 0.204082) (8., 0.1875) (9., 0.172839) (10,0.16)%
\psdots[linecolor=c2,dotstyle=diamond*] (3., 0.888889) (4., 1.) (5., 0.96) (6., 0.888889) (7., 0.816327) (8., 0.75) (9., 0.691358) (10., 0.64)%
\psdots[linecolor=c7,dotstyle=square*](1.83658, 0.239496) (2.93114, 0.262076) (3.94516, 0.241034) (4.95104, 0.214254) (5.95429, 0.19046) (6.95634, 0.17055) (7.95777, 0.154021) (8.95881, 0.14022) (9.9596, 0.128583)%
\psdots[linecolor=c4,dotstyle=triangle*](2.20969, 0.665598) (3.12783, 0.924951) (4.1036, 0.893546) (5.09255, 0.812039) (6.08628, 0.7309) (7.08225, 0.65972) (8.07945, 0.599081) (9.07739, 0.547613)%
\end{pspicture}
\begin{pspicture}(-1,-0.3)(10,1.2)
\psline(-0.2,0)(10,0) \rput(-0.5,0){$0$}%
\psline(-0.15,0.2)(0,0.2)%
\psline(-0.2,0.4)(0,0.4) \rput(-0.5,0.4){$0.1$}%
\psline(-0.15,0.6)(0,0.6)%
\psline(-0.2,0.8)(0,0.8) \rput(-0.5,0.8){$0.2$}%
\psline(-0.15,1)(0,1)%
\psline(-0.2,1.2)(0,1.2) \rput(-0.5,1.2){$0.3$}%
\rput{90}(-1.4,1.0){${\cal N}_{1m1}/(\epsilon \varpi T)^2$}
\psline(0,-0.066666666)(0,1.2) \rput(0,-0.166666666666){$0$}%
\psline(1,-0.05)(1,0)%
\psline(2,-0.05)(2,0)%
\psline(3,-0.05)(3,0)%
\psline(4,-0.05)(4,0)%
\psline(5,-0.066666666)(5,0) \rput(5,-0.166666666666){$5$}%
\psline(6,-0.05)(6,0)%
\psline(7,-0.05)(7,0)%
\psline(8,-0.05)(8,0)%
\psline(9,-0.05)(9,0)%
\psline(10,-0.066666666)(10,0) \rput(10,-0.16666666666){$10$}%
\rput(8,-0.333333333333){${\varpi} \pi/(r_o-r_i)$}%
\psdots (2.09194, 0.193801) (3.07064, 0.190726) (4.06265, 0.165563) (5.05855, 0.143062) (6.05606, 0.125047) (7.05439, 0.110725) (8.05319, 0.0991963) (9.05229, 0.0897666)%
\psdots[linecolor=c2,dotstyle=diamond*] (2.09194, 0.88403) (3.07064, 0.829703) (4.06265, 0.7131) (5.05855, 0.613977) (6.05606, 0.53576) (7.05439, 0.473962) (8.05319, 0.424377) (9.05229, 0.383897)%
\psdots[linecolor=c7,dotstyle=square*] (0.913954, 0.197624) (1.95549, 0.152794) (2.95665, 0.11437) (3.95686, 0.0900872) (4.95692, 0.0740416) (5.95695, 0.0627642) (6.95696, 0.0544353) (7.95696, 0.048043) (8.95697, 0.0429866) (9.95697, 0.038889)%
\psdots[linecolor=c4,dotstyle=triangle*]  (1.46733, 0.490273) (2.33204, 0.561268) (3.29386, 0.471634) (4.27688, 0.39238) (5.26734, 0.332865) (6.26125, 0.288043) (7.25702, 0.253464) (8.25391, 0.226113) (9.25153, 0.203994)%
\end{pspicture}
\end{center}
\caption{Plot of $\mathcal{N}_{lms}/(\epsilon\varpi T)^{2}$ as a
function of $\varpi \pi/(r_{o}-r_{i})$, in the resonance condition
for a few values of $l$ and $s$. We have considered
Dirichlet-Dirichlet and Mixed BCs. On the legend (top-right of the
figure), the letter on the left indicates the BC imposed on the
field at the inner shell and the  letter to the right, indicates the
BC imposed on the field at the outer shell. $D$ means Dirichlet BC
and static shell, $\tilde{D}$ means Dirichlet BC and moving shell,
whereas $N$ means Neumann BC and static shell. We have set
$r_{o}=2r_{i}$.}
\end{figure}

In the limit $r_{i}\gg r_{o}-r_{i}$, we can use the Bessel
asymptotic forms for large arguments to derive an analytical
expression for the average number of created particles in a
particular mode. For the case where the field is submitted to
Dirichlet-Dirichlet BCs, we obtain
\begin{equation}
\omega_{lss^{\prime}}
 \;\;\;\longrightarrow\;\;\;
 \frac{(s+s^{\prime})\pi}{\left\vert r_{\alpha
}-r_{\beta}\right\vert }\label{19}%
\end{equation}
and
\begin{equation}
\lim_{\omega\rightarrow\omega_{lss^{\prime}}}\mathcal{N}_{lms}
 \;\;\;\longrightarrow\;\;\;
\frac{\epsilon^{2}\pi^{2}T^{2}}{4}\frac{r_{\beta}^{2}}{\left(  r_{\alpha
}-r_{\beta}\right)  ^{4}}s^{\prime}s.\label{20}%
\end{equation}
Analogously, for the field submitted to mixed BCs (\ref{11}), we
have
\begin{equation}
\omega_{lss^{\prime}}\rightarrow\frac{(s+s^{\prime}-1)\pi}{\left\vert
r_{\alpha}-r_{\beta}\right\vert }\label{21}%
\end{equation}
and
\begin{equation}
\lim_{\omega\rightarrow\omega_{lss^{\prime}}}\mathcal{N}_{lms}\rightarrow
\frac{\epsilon^{2}\pi^{2}T^{2}}{16}\frac{r_{\beta}^{2}}{\left(  r_{\alpha
}-r_{\beta}\right)  ^{4}}(2s^{\prime}-1)(2s-1).\label{22}%
\end{equation}
Expressions (\ref{19}) and (\ref{20}) correspond to the results for
the $1+1$ DCE under Dirichlet-Dirichlet BCs derived in Refs.
\cite{DodonovKlimov1996,JungSoh1998,DodonovRevisao2001}, whereas
Eqs. (\ref{21}) and (\ref{22}) correspond to the results under mixed
BCs presented in
\cite{Hushwater1997,Cougo-PintoEtAl1999,AguiarEtAl2003}. These
similarities can be related to the fact that the limit $r_{i} \gg
r_{o}-r_{i}$ is akin to the plane geometry.

\section{Concluding remarks}

In this paper we have investigated the dynamical Casimir effect for
a massless scalar field within two concentric spherical shells
considereing mixed boundary conditions. We have thus complemented
some previous results presented in Ref. \cite{PascoalEtAl2008} where
the massless scalar field was assumed to satisfy only  Dirichlet BC
in both shells. We have analyzed the real particle creation
phenomenon for the case where only one of the shells is allowed to
move with an arbitrary  law of motion for its radius. In addition,
the Dirichlet BC was imposed on the moving shell while the Neumann's
was assumed on the static one. However, in our discussion, the
moving shell could be  the inner shell or the outer one as well. In
order to get some numerical estimatives, and with the purpose of
comparing our results with those obtained in Ref.
(\cite{PascoalEtAl2008}), we chose a particular, but very typical,
oscillating motion for the moving shell, in which it starts moving
at a certain instant, oscillates with a given frequency and then
stops suddenly its motion. Considering this  particular situation,
we have identified the resonance conditions where the number of
created particles is more appreciable. A direct inspection in our
graphs (see Fig. (4)), allows us to make some conclusions: for both
cases of Dirichlet-Dirichlet BCs  or mixed BCs, we see that every
time the moving shell is the outer one the average number of created
particles is greater than the corresponding cases where the inner
shell is in motion (by a factor of the order of $\sim 4$) . This can
be unsderstood simply recalling that the dynamical Casimir effect
increases with the area of the moving surface. In other words, the
dissipative force that acts on the moving boundary, responsible for
converting mechanical energy into field energy (real field {\it
quanta}) increases with the area with the moving boundary. Another
interesting result that can be extracted from our calculations is
the fact that the case with mixed BCs presents lower resonance
frequencies than that with Dirichlet-Dirichlet BCs. This feature can
be useful for further experimental investigations of particle
creation within the context of the dynamical Casimir effect, since
it makes easier to access the parametric amplification regime of
particle creation.

\end{document}